\documentclass{article}

\usepackage{graphicx}
\usepackage{axodraw}

\newcommand{\be}{\begin{equation}}
\newcommand{\ee}{\end{equation}}
\newcommand{\bea}{\begin{eqnarray}}
\newcommand{\eea}{\end{eqnarray}}
\begin{document}

\begin{titlepage}

\begin{flushright}
Edinburgh 2002/16\\
IPPP/02/56\\
DCPT/02/112\\
October 2002
\end{flushright}
\vspace{1.cm}

\begin{center}
\large\bf
{\LARGE\bf A numerical evaluation of the scalar hexagon integral in the physical
region}\\[1cm]
\rm
{T.~Binoth$^{a}$, G.~Heinrich$^{b}$ and N.~Kauer$^{a}$}\\[1cm]

{\em $^{a}$Department of Physics and Astronomy,\\
           The University of Edinburgh,
	   EH9 3JZ Edinburgh, Scotland} \\[.5cm]

{\em $^{b}$IPPP, University of Durham, Durham DH1 3LE, UK} \\[.5cm]

\end{center}
\normalsize

\vspace{1cm}

\begin{abstract}
We derive an analytic expression for the scalar
one-loop pentagon and hexagon functions 
which is convenient for subsequent numerical integration. 
These functions are of relevance in the computation of
next-to-leading order radiative corrections 
to multi-particle cross sections. The 
hexagon integral is represented in terms of $n$-dimensional
triangle functions and ($n$+2)-dimensional box functions.
If infrared poles are present this representation
naturally splits into a finite and a pole part.
For a fast numerical integration of the finite part 
we propose simple one- and two-dimensional integral representations.
We set up an iterative numerical integration method to calculate
these integrals directly in an efficient way. 
The method is  illustrated by explicit results for pentagon and
hexagon functions with some generic physical kinematics.    
\end{abstract}



\end{titlepage}

\section{Introduction}

The growing potential of current and future high energy colliders
permits an increasingly detailed and precise study of particle physics and
will push the energy frontier to the TeV scale before the end of this
decade.  Experiments at the Fermilab Tevatron, the LHC at CERN and the
planned $e^+e^-$ linear collider will produce vast amounts of data.
The prospect of unprecedented experimental statistics calls for a 
corresponding level of precision on the theoretical side which 
generally requires higher-order predictions for the underlying 
multi-particle interactions.
To achieve this objective is particularly important for the discovery of
physics beyond the Standard Model, where a precise description of Standard
Model backgrounds is crucial and necessitates next-to-leading order (NLO)
calculations for a variety of multi-particle search channels.
The resulting reduction of unphysical scale uncertainties
allows for precise quantitative predictions, thus significantly improving
the rather qualitative leading order descriptions.
For background predictions the importance of higher-order effects is further
enhanced by severe experimental cuts, 
because the latter typically select peripheral phase
space regions, which are particularly sensitive to NLO effects. 
 
While for partonic and electroweak $2\rightarrow 2$ processes advanced theoretical
knowledge about NLO calculations exists,
our understanding of processes with more final-state particles
has not yet reached maturity.
In recent years, many relevant $2\rightarrow 3$ amplitudes and processes involving
a small number of scales (as typical in QCD) have been calculated at NLO.
However, results for multi--scale $2\rightarrow 3$ processes are rare, and
the step to $2\rightarrow 4$ amplitudes again entails a significant 
increase in complexity.
Currently, $2\to 4$ results are only known for certain supersymmetric gauge theory
amplitudes~\cite{Bern:1994cg} and the Yukawa model \cite{Binoth:2001vm}. 

NLO calculations for a $2\rightarrow N$ process 
are typically organized by evaluating the real emission 
$2\rightarrow N+1$ part and the virtual corrections 
to the $2\rightarrow N$ process separately\footnote{An exception is
the purely numerical approach of Soper \cite{soper}, which
combines real and virtual corrections on the integrand level.}. 
In the latter, divergences are regulated by working in $n=4-2\epsilon$
dimensions. The analytic evaluation of the virtual part requires the reduction 
of tensor integrals to scalar integrals, which is well understood 
for arbitrary $2\rightarrow N$ processes \cite{Bern:1995ix,Binoth:1999sp,Tarasov:1996br}. 
Reduction formulas for scalar integrals
which relate general finite $N$-point functions to box integrals
have been formulated in \cite{VanNeervenVermaseren}
for the infrared finite case.
For massless integrals with $N\leq $~7, similar reduction formulas have been
derived in $n$ dimensions in \cite{BernDixonKosower}. 
The generalization to arbitrary $N$ has been considered 
in \cite{Tarasov:1996br,Fleischer:1999hq}
for massive integrals and 
in \cite{Binoth:1999sp} for the massless case. 

As will be pointed out below,
the reduction approach leads to a natural separation between
IR divergent and finite contributions. The IR divergences can be 
collected in  triangle functions with massless propagators 
which  have simple analytic representations.
The finite remainder of the scalar integrals
can be expressed fully analytically in terms of  
dilogarithms (and simpler functions)~\cite{tHooftVeltman}.
However, the large number of kinematic invariants leads to a huge number of 
dilogarithms,
many thousands in the case of the massive hexagon function.
A numerical evaluation at this level typically leads to
large cancellations  in certain 
kinematic regions  and thus to numerical instabilities \cite{vanOldenborgh:1989wn}.
Hence a numerical evaluation at an earlier stage may be 
equally good for practical purposes, if not better, and this is what we 
suggest in this paper. 

An alternative numerical approach to the evaluation of Feynman diagrams
which deals with the computation of multi-leg one-loop diagrams
has been presented in \cite{Passarino:2001wv,Ferroglia:2002mz}. 
It is based on the Bernstein-Tkachov theorem \cite{Tkachov:1996wh}. 
The basic idea is to rise the power of negative exponents of kinematic functions
by using the Bernstein-Tkachov relation. This leads in principle
to better behaved integrands. 
However, an explicit result for the hexagon function has not been given yet.

In this paper we derive a representation of the hexagon function
in terms of 20 $n$-dimensional triangle functions and 15 ($n$+2)-dimensional
box functions. For completeness we also provide the pentagon 
function in terms of 10 triangle and 6 box functions.
By using sector decomposition and one explicit
integration we derive a one-dimensional integral representation
for the triangle and a two-dimensional parameter integral for the box.
We study the analytic structure of these representations in some detail. 
The virtue of these parameter integral representations lies in the fact 
that the singularity structure is quite transparent. Hence, 
as a by-product, we derive the well-known fact that for
physical kinematics only integrable singularities
of logarithmic and square-root type are present at one loop.
 
Further, we describe a numerical  integrator 
that facilitates a fast and accurate evaluation of these ``atoms'' 
of our representation.      
Finally we give numerical examples for our approach.
We only consider the case where all propagators are
massive here, but  also comment on
how to generalize our approach in the presence of IR divergences.

The paper is organized as follows. In Section 2 we
derive representations for the $n$-dimensional box, pentagon and
hexagon functions in terms of $n$-dimensional triangle
and ($n$+2)-dimensional box functions.  In Section 3,
one- and two-dimensional integral representations
are given for the triangles and boxes, together with a detailed
discussion of the singularity structure of the integrands.
In Section 4 we outline the numerical evaluation of these
integral representations. Examples for our procedure
are provided in Section 5. The article closes with
a summary and an outlook.   

\section{Reduction of pentagon and hexagon integrals}

In this section we will provide explicit expressions
for the $n$-dimensional hexagon, pentagon and box functions
in terms of $n$-dimensional triangle and ($n$+2)-dimensional
box functions. If infrared divergences are present, which
manifest themselves in terms of poles  in $ 1/\epsilon=2/(4-n)$,
this choice of building blocks naturally splits these functions into
infrared divergent and finite pieces. After separating
the IR divergent triangles from the representation, only 
integrals which can be evaluated in $n=4$ dimensions remain.
To fix our conventions we define the $n$-dimensional 
$N$-point function as
\begin{eqnarray}
I_N^n(p_1,\dots,p_N,m_{1},\dots,m_{N}) &=& 
\int  \frac{d^nk}{i\pi^{n/2}}\; \frac{1}{\prod^N_{l=1} [(k-r_l)^2-m_l^2]} \nonumber\\
&=& 
(-1)^N \Gamma(N-n/2) \int_{0}^{\infty} d^Nx 
\frac{\delta(1-\sum_{l=1}^N x_l)}{(x\cdot S \cdot x/2)^{N-n/2}} 
\label{EQdefi}
\end{eqnarray}
The kinematic information is contained in the matrix $S$
which is related to the Gram matrix $G$ by
\begin{eqnarray} \label{EQStoG}
S_{kl}&=&-(r_l-r_k)^2 + m_l^2 + m_k^2 = ( G_{kl} - v_l -v_k )\\
G_{kl}&=&2\,r_k\cdot r_l\; ,\quad v_k=G_{kk}/2 - m_k^2 \; ,
\quad k,l=1,\ldots,N \nonumber
\end{eqnarray}
The vectors $r_j=p_1+\dots +p_j$ are sums of external momenta. 
In physical applications the external momenta span the 4-dimensional
Minkowski space. We consider here only the case where any four
of the $N$ external vectors are linearly independent. 
Further we assume that the kinematics is such that the anomalous
threshold, i.e. the leading singularity of the $N$-point
function, is not probed. We recall that the leading singularity of a Feynman integral
in parameter space corresponds to  
a vanishing  denominator of the parameter integral while 
 {\em all} values of the Feynman parameters are nonzero~\cite{analyticsmatrix}. 
This means that the matrix $S$ has to have
a zero eigenvalue or simply that $\det(S)=0$. 
The generalization of our approach to this exceptional case is briefly discussed 
below, but let us first focus on non-exceptional kinematic configurations.
In this case
every $N$-point function with $N\geq 6$ can be expressed
in terms of pentagon integrals~\cite{Binoth:1999sp}. 
The pentagon functions, in turn, can be
expressed in terms of box functions up to a term which
vanishes in the limit $n\rightarrow 4$.
For the purpose of this paper, we only give the reduction formula 
for the case $N\le 6$.
It is derived in the same way as in \cite{Binoth:1999sp}, where a simple 
derivation for the massless case and general $N$ is explicitly given. 
\begin{eqnarray}\label{EQscalarreductionNT}
I_N^n &=& \sum \limits_{k=1}^{N} B_k \,
  I_{N-1,k}^n  + (N-n-1) \,\frac{\det(G)}{\det(S)}\, I_N^{n+2} 
  \quad , \quad \det(S)\not =0, \\
  B_k &=&  -\sum\limits_{l=1}^{N} S^{-1}_{kl} 
\end{eqnarray}
Hence the $N$-point function decays into a sum of ($N$--1)-point functions,
which are obtained by pinching propagators
\begin{eqnarray}\label{EQreducedgraph}
I^n_{N-1,j} =\int \frac{d^nk}{i\pi^{n/2}} \; 
\frac{ [(k-r_j)^2 - m_j^2] }{\prod^N_{l=1}[(k-r_l)^2 - m_l^2] }\;,
\end{eqnarray}
and a remainder term which is proportional to the Gram determinant,
$\det(G)$,
times the ($n$+2)-dimensional $N$-point function. The latter
turns out to be infrared finite even in the massless case, 
as can be seen by power counting.
The reduction coefficients $B_k$ are defined through the inverse of 
the kinematic matrix $S$.
Note that the formula is valid for general dimension, as long as 
the external momenta are non-exceptional.
The rank of the Gram matrix is
$\min(4,N-1)$, whereas the rank of the matrix $S$ is $\min(6,N)$, 
such that the inverse of $S$ does not exist for $N>6$.  
However, we stress that reduction formulas do not rely on the
regularity of the matrix $S$, since 
in the case of a singular $S$ one can follow the lines of  
\cite{Binoth:1999sp}, where the concept of a pseudo-inverse to a matrix
was used to deal with the case  $N>6$. 
The case $N\le 6$ with $\det(S)=0$ can be treated analogously.

In the generic case of non-exceptional momenta the hexagon function decays into 
six pentagons without rest
\begin{eqnarray}\label{EQSR6}
&&I_6^n(s_{12},s_{23},s_{34},s_{45},s_{56},s_{61},s_{123},s_{234},s_{345},
s_{1},\dots,s_{6},m_{1},\dots,m_{6}) =  \nonumber\\
&&\hspace{2cm}\quad B_1 \, I_5^n(s_{123},s_{34},s_{45},s_{56},s_{345},s_{12},s_{3},s_{4},s_{5},s_{6},m_2,m_3,m_4,m_5,m_6)\nonumber\\
&&\hspace{2cm}     +B_2 \, I_5^n(s_{234},s_{45},s_{56},s_{61},s_{123},s_{23},s_{4},s_{5},s_{6},s_{1},m_3,m_4,m_5,m_6,m_1)\nonumber\\
&&\hspace{2cm}     +B_3 \, I_5^n(s_{345},s_{56},s_{61},s_{12},s_{234},s_{34},s_{5},s_{6},s_{1},s_{2},m_4,m_5,m_6,m_1,m_2)\nonumber\\
&&\hspace{2cm}     +B_4 \, I_5^n(s_{123},s_{61},s_{12},s_{23},s_{345},s_{45},s_{6},s_{1},s_{2},s_{3},m_5,m_6,m_1,m_2,m_3)\nonumber\\
&&\hspace{2cm}     +B_5 \, I_5^n(s_{234},s_{12},s_{23},s_{34},s_{123},s_{56},s_{1},s_{2},s_{3},s_{4},m_6,m_1,m_2,m_3,m_4)\nonumber\\
&&\hspace{2cm}     +B_6 \, I_5^n(s_{345},s_{23},s_{34},s_{45},s_{234},s_{61},s_{2},s_{3},s_{4},s_{5},m_1,m_2,m_3,m_4,m_5)\\
&&\hspace{2cm} B_k =  -\sum\limits_{l=1}^{6} (S^{(6)}_{kl})^{-1} \nonumber
\end{eqnarray}
The kinematic matrix is 
\begin{eqnarray*}
S_{kl}^{(6)} &=& \hat S_{kl}^{(6)} + m_k^2 + m_l^2 \qquad (k,l=1,\ldots, 6)\\
\hat S^{(6)} &=& -\left(  \begin{array}{llllll} 
 0      &  s_2     &  s_{23} & s_{234}& s_{16}  & s_1    \\
s_2    &    0      &  s_3    & s_{34} & s_{345} & s_{12}  \\ 
s_{23} &  s_3     &    0     & s_4    & s_{45}  & s_{123} \\
s_{234}&  s_{34}  &  s_4     &    0   & s_5     & s_{56}    \\
s_{61} &  s_{345} &  s_{45}  & s_5    &   0     & s_6 \\
s_1    &  s_{12}  &  s_{123} & s_{56} & s_6     & 0  
\end{array}  \right)
\end{eqnarray*}
The kinematic invariants are defined as 
$s_j=p_j^2$,
$s_{ij\ldots}=(p_i+p_j+\ldots\quad)^2$.\\
Similarly, the general pentagon integral can be written as
\begin{eqnarray}\label{EQSR5}
&&I_5^n(s_{12},s_{23},s_{34},s_{45},s_{51},s_{1},\dots,s_{5},m_{1},\dots,m_{5}) =  \nonumber\\
&&\hspace{3cm}\quad B_1 \,  I_4^n(s_{45},s_{34},s_{12},s_{3},s_{4},s_{5},m_2,m_3,m_4,m_5)\nonumber\\
&&\hspace{3cm}+B_2 \,  I_4^n(s_{51},s_{45},s_{23},s_{4},s_{5},s_{1},m_3,m_4,m_5,m_1)\nonumber\\
&&\hspace{3cm}+B_3 \,  I_4^n(s_{12},s_{51},s_{34},s_{5},s_{1},s_{2},m_4,m_5,m_1,m_2)\nonumber\\
&&\hspace{3cm}+B_4 \,  I_4^n(s_{23},s_{12},s_{45},s_{1},s_{2},s_{3},m_5,m_1,m_2,m_3)\nonumber\\
&&\hspace{3cm}+B_5 \,  I_4^n(s_{34},s_{23},s_{51},s_{2},s_{3},s_{4},m_1,m_2,m_3,m_4)+ {\cal O}(\epsilon)\\
&& \mbox{here } \quad B_k =  -\sum\limits_{l=1}^{5} (S^{(5)}_{kl})^{-1}\qquad,\quad 
S_{kl}^{(5)} = \hat S_{kl}^{(5)} + m_k^2 + m_l^2 \qquad (k,l=1,\ldots,5)\nonumber\\
&&\hspace{3cm}\hat S^{(5)} = -\left(  \begin{array}{lllll} 
0       & s_2     &  s_{23} & s_{51} & s_1     \\
s_2    &  0       &  s_3    & s_{34} & s_{12}  \\ 
s_{23} & s_3     &  0       & s_4    & s_{45}  \\
s_{51} & s_{34}  & s_4   & 0       & s_5     \\
s_1    & s_{12}  & s_{45}  & s_5    & 0   
\end{array}  \right) \nonumber
\end{eqnarray}
In principle, 
if no infrared divergences are present, there is no need to
reduce any further, as analytic
formulas for the finite 4-dimensional box integral exist. The boxes can be
expressed in terms of a large number of dilogarithms
\cite{Denner:qq}. However, as is well known \cite{vanOldenborgh:1989wn},
these representations are not unproblematic from a numerical point of view,
since large cancellations between the dilogarithms occur in certain kinematic
regimes.
For this reason, and also in view of the general case which includes 
infrared divergences,
it will turn out to be useful to reduce the box integrals further. 
This allows for a natural
separation of the IR-singular and finite terms. 

The reduction formula for boxes reads
\begin{eqnarray}\label{box}
&& I_4^n(s_{12},s_{23},s_{1},\dots,s_{4},m_{1},\dots,m_{4}) = \nonumber\\
&&\hspace{3cm}\quad B_1 \,  I_3^n(s_{12},s_{3},s_{4},m_2,m_3,m_4)\nonumber\\
&&\hspace{3cm}+B_2      \,  I_3^n(s_{23},s_{4},s_{1},m_3,m_4,m_1)\nonumber\\
&&\hspace{3cm}+B_3      \,  I_3^n(s_{12},s_{1},s_{2},m_4,m_1,m_2)\nonumber\\
&&\hspace{3cm}+B_4      \,  I_3^n(s_{23},s_{2},s_{3},m_1,m_2,m_3)\nonumber\\
&&\hspace{3cm}+(n-3) ( B_1+B_2+B_3+B_4 ) \,
I_4^{n+2}(s_{12},s_{23},s_{1},\dots,s_{4},m_{1},\dots,m_{4})\\
&& B_k =  -\sum\limits_{l=1}^{4} (S^{(4)}_{kl})^{-1}\qquad,\quad 
S_{kl}^{(4)} = \hat S_{kl}^{(4)} + m_k^2 + m_l^2 \qquad (k,l=1,\ldots, 4)\nonumber\\
&&\hat S^{(4)} = -\,\left(  \begin{array}{cccc} 
0     &  s_2        &  s_{23}     &  s_1 \\
s_2   &    0        &  s_3        &  s_{12}     \\ 
s_{23}&  s_3        &  0          &  s_4 \\
s_1   &    s_{12}   &  s_4        &  0  
\end{array}  \right)\nonumber
\end{eqnarray}
Note that $\det(G^{(4)})=-(B_1+B_2+B_3+B_4)\det(S^{(4)})$.

By applying the reduction formulas iteratively
one finds a representation of the pentagon integral in terms of 
10 triangle and 5 ($n$+2)-dimensional box integrals.
It can be written in the following
compact, cyclically symmetric form
\begin{eqnarray}\label{pentagon}
I_5^n &=&\quad ( B_1 B_{12} + B_2 B_{21} ) I_{3,12}^n 
         +( B_1 B_{13} + B_3 B_{31} ) I_{3,13}^n 
         + B_1 ( B_{12} + B_{13} + B_{14} + B_{15} )  I_{4,1}^{n+2}\nonumber\\
&& + ( B_2 B_{23} + B_3 B_{32} ) I_{3,23}^n 
   + ( B_2 B_{24} + B_4 B_{42} ) I_{3,24}^n 
   +   B_2 ( B_{21} + B_{23} + B_{24} + B_{25} )  I_{4,2}^{n+2}	 \nonumber\\
&& + ( B_3 B_{34} + B_4 B_{43} ) I_{3,34}^n 
   + ( B_3 B_{35} + B_5 B_{53} ) I_{3,35}^n 
   +   B_3 ( B_{31} + B_{32} + B_{34} + B_{35} )  I_{4,3}^{n+2}\nonumber\\
&& + ( B_4 B_{45} + B_5 B_{54} ) I_{3,45}^n 
   + ( B_4 B_{41} + B_1 B_{14} ) I_{3,14}^n 
   +   B_4 ( B_{41} + B_{42} + B_{43} + B_{45} )  I_{4,4}^{n+2}\nonumber\\
&& + ( B_5 B_{51} + B_1 B_{15} ) I_{3,15}^n 
   + ( B_5 B_{52} + B_2 B_{25} ) I_{3,25}^n 
   +   B_5 ( B_{51} + B_{52} + B_{53} + B_{54} )  I_{4,5}^{n+2}    
 \nonumber\\ && 
\end{eqnarray}
Here the $B_i$ are the reduction coefficients of the pentagon integral (\ref{EQSR5})
and $B_{ij}$ is the $j$th reduction coefficient of
that box integral which stems from the  $i$th pinch of
the pentagon integral. Note that $B_{ij}\neq B_{ji}$.
On the other hand, the triangles, which result from double pinches 
of the pentagons, are symmetric:
$I_{3,ij}^n=I_{3,ji}^n$. 
Analogously, one finds for the hexagon integral a representation
in terms of 20 triangle and 15 ($n$+2)-dimensional box integrals:
\begin{eqnarray}\label{hexagon}
I_6^n = \hspace{13cm}\nonumber\\
\qquad \Bigl\{ \left[ B_1 ( B_{12}  B_{123} + B_{13} B_{132}  ) +
               B_2 ( B_{21}  B_{123} + B_{23} B_{231}  ) +
               B_3 ( B_{31}  B_{132} + B_{32} B_{231}  ) \right] I_{3,123}^n 
+ 5 \;\mbox{c.p.}\Bigr\}\hspace{.2cm} \nonumber\\  +
\Bigl\{ \left[ B_1 ( B_{12}  B_{124} + B_{14} B_{142}  ) +
                 B_2 ( B_{21}  B_{214} + B_{24} B_{241}  ) +
		 B_4 ( B_{41}  B_{412} + B_{42} B_{421}  ) \right] I_{3,124}^n 
 + 5 \;\mbox{c.p.} \Bigr\}\nonumber\\ +
\Bigl\{ \left[ B_1 ( B_{13}  B_{134} + B_{14} B_{143}  ) +
                 B_3 ( B_{31}  B_{314} + B_{34} B_{341}  ) +
		 B_4 ( B_{41}  B_{413} + B_{43} B_{431}  ) \right] I_{3,134}^n 
 + 5 \;\mbox{c.p.} \Bigr\}\nonumber\\  +
\Bigl\{ \left[ B_1 ( B_{13}  B_{135} + B_{15} B_{153}  ) +
                 B_3 ( B_{31}  B_{315} + B_{35} B_{351}  ) +
		 B_5 ( B_{51}  B_{513} + B_{53} B_{531}  ) \right] I_{3,135}^n 
+ 1\; \mbox{c.p.} \Bigr\}\nonumber\\ 
 +\Bigl\{( B_1 B_{12} + B_2 B_{21} ) ( B_{123}+ B_{124}+B_{125}+B_{126} )
I_{4,12}^{n+2} + 5 \;\mbox{c. p.}\Bigr\}\nonumber\\  +
\Bigl\{( B_1 B_{13} + B_3 B_{31} ) 
( B_{132}+ B_{134}+B_{135}+B_{136} )I_{4,13}^{n+2} 
+ 5 \;\mbox{c.p.}\Bigr\}\nonumber\\ 
+\Bigl\{( B_1 B_{14} + B_4 B_{41} ) 
( B_{142}+ B_{143}+B_{145}+B_{146} )I_{4,14}^{n+2}
+ 2\; \mbox{c.p.}\Bigr\}\nonumber\\ 
\end{eqnarray}
Here the $B_i$ are the reduction coefficients of the 
hexagon integral (\ref{EQSR6}).
The $B_{ijk}=B_{jik}$ are a shorthand for the reduction 
coefficient of the $k$th pinch of the box integral $I^n_{4,ij}$ 
and c.p. means cyclic permutation of the indices of the $B$'s
and the indices which define the pinches. 
The problem of calculating the pentagon and hexagon
integrals is now reduced to the calculation of 
lower point integrals and reduction coefficients. 
As pointed out above, a complete 
analytic expression is possible but not necessarily
of advantage.
Such an analytic expression contains a huge number of dilogarithms,
and nontrivial numerical cancellations occur during evaluation.
Since one has to rely on numerical integration at some stage
of the calculation of most physical cross sections anyhow, 
a direct numerical evaluation of the scalar integrals in parameter form
is more than adequate for practical applications.
In our approach, all one has to do is to provide stable and sufficiently
fast numerical integrators for the finite 4-dimensional
triangle and 6--dimensional box integrals.

Numerical instabilities typically arise from terms with opposite signs and
denominators that approach zero.
The denominators that occur in our reduction are the determinants
of the kinematic matrices $S$ from the different reduced
hexagon, pentagon and box integrals.
Thus, the critical points are the normal and anomalous
thresholds~\cite{analyticsmatrix} 
of the corresponding scalar graph. Near these thresholds one finds that 
the reduction coefficients fulfill (approximately) additional constraints. 
This can be exploited to achieve stable groupings of terms in the respective
limits. 
  
\section{Integral representations of triangle and box functions}

In this section we first derive integral representations of the triangle and 
box functions which are appropriate for direct numerical integration.
Then we thoroughly analyse the singularity structure of the integrands.

Using standard Feynman parametrisation, the parameter representations 
of the $n$-dimensional 3- and 4-point functions are
\bea
I_3^{n}(s_1,s_2,s_3,m_1^2,m_2^2,m_3^2) &=& 
\int \frac{d^nk}{i\pi^{n/2}} 
\frac{1}{[(k-r_1)^2-m_1^2][(k-r_2)^2-m_2^2][k^2-m_3^2]} \nonumber \\
&=& -\Gamma(3-n/2)\int\limits_0^\infty d^3x \, \delta(1-x_{123})
\frac{(x_{123})^{3-n}}{\left({\cal F}_{Tri}\right)^{3-n/2}} 
\eea 
\bea
{\cal F}_{Tri} &=&(-s_1) x_1x_3 + (-s_2) x_1x_2+ (-s_3) x_2x_3
+x_{123}(x_1 m_1^2 + x_2 m_2^2 + x_3 m_3^2) - i\delta\nonumber\\
x_{123}&=&x_1+x_2+x_3\nonumber
\eea
and 
\bea
&& I_4^{\hat{n}}(s_{12},s_{23},s_1,s_2,s_3,s_4,m_1^2,m_2^2,m_3^2,m_4^2) \hspace{4cm}\nonumber \\
&&\qquad\qquad=\int \frac{d^{\hat{n}}k}{i\pi^{\hat{n}/2}} 
\frac{1}{[(k-r_1)^2-m_1^2][(k-r_2)^2-m_2^2][(k-r_3)^2-m_3^2][k^2-m_4^2]} \nonumber \\
&&\qquad\qquad= \Gamma(4-\hat{n}/2)\int\limits_0^\infty d^4x \, \delta(1-x_{1234})
\frac{(x_{1234})^{4-\hat{n}}}{\left({\cal F}_{Box}\right)^{4-\hat{n}/2}}
\eea 
\bea
{\cal F}_{Box} &=& (-s_{12}) x_2 x_4 + (-s_{23}) x_1 x_3
                      + (-s_1)  x_1 x_4 + (-s_2)  x_1 x_2
		      + (-s_3)  x_2 x_3 + (-s_4)  x_3 x_4 \nonumber \\
&&+x_{1234}(x_1 m_1^2 + x_2 m_2^2 + x_3 m_3^2+ x_4 m_4^2)
\eea 
Note that we will need the case $\hat{n}=6-2\epsilon$ for the box in the following.

In special cases, e.g. if the integral has a high symmetry
due to equal scales or vanishing kinematic invariants,
a non-symmetric transformation of Feynman parameters
typically leads to simplifications. However, here we want to deal with
general situations, so no Feynman parameters should be preferred
and a symmetric treatment of the problem seems adequate.    
This is achieved by splitting the 
$N$-dimensional ($N=3,4$) parameter integrals 
into $N$ sectors by the following decomposition
\bea\label{sectordeco}
1 = \theta(x_1>x_2,\dots,x_N) +\theta(x_2>x_1,x_3,\dots,x_N) 
+ \dots + \theta(x_N>x_1,\dots,x_{N-1})
\eea
This approach will turn
out to be useful for numerical integration later. In addition, it is very 
convenient in the presence of IR divergences~\cite{Binoth:2000ps}. 
The box integral then decays into a sum of four 3-dimensional parameter integrals,
while the triangle integral decays into a sum of three 2-dimensional parameter
integrals. 
Focusing on the last sector as an explicit example, one finds
after the variable transformation $t_j=x_j/x_N$  $(j=1,\ldots,N-1)$
\bea\label{Stri}
S_{Tri}^n(s_1,s_2,s_3,m_1^2,m_2^2,m_3^2) 
= \int\limits_0^1 dt_1 dt_2\, 
\frac{(1+t_{1}+t_2)^{3-n}}{\left(\tilde{\cal F}_{Tri}\right)^{3-n/2}} 
\eea 
\bea
\tilde{\cal F}_{Tri} =(-s_1) t_1 + (-s_2) t_1t_2+ (-s_3) t_2
+(1+t_{1}+t_2)(t_1 m_1^2 + t_2 m_2^2 + m_3^2) 
\eea 
\bea\label{Sbox}
S_{Box}^{n+2}(s_{12},s_{23},s_1,s_2,s_3,s_4,m_1^2,m_2^2,m_3^2,m_4^2) 
&=& \int\limits_0^1 dt_1 dt_2 dt_3\,  
\frac{(1+t_{1}+t_2+t_3)^{2-n}}{\left(\tilde{\cal F}_{Box}\right)^{3-n/2}}
\eea 
\bea
\tilde{\cal F}_{Box} &=& (-s_{12}) t_2  + (-s_{23}) t_1 t_3
                      + (-s_1)  t_1 + (-s_2)  t_1 t_2
		      + (-s_3)  t_2 t_3 + (-s_4)  t_3 \nonumber \\
&&+(1+t_{1}+t_2+t_3)(t_1 m_1^2 + t_2 m_2^2 + t_3 m_3^2+ m_4^2)
\eea 
Since in the Euclidean region both functions $\tilde{\cal F}_{Box}$ and 
$\tilde{\cal F}_{Tri}$ are strictly positive, bounded functions, 
it is clear that the respective integrals can 
easily be computed numerically. We will see shortly that the above
representations allow for a transparent discussion of the singularity structure 
of the integrals for physical kinematics. 
The box and triangle functions are given in terms of the sector functions
(\ref{Stri}) and (\ref{Sbox}) 
as
\bea
&&I_3^{n}(s_1,s_2,s_3,m_1^2,m_2^2,m_3^2) = 
-\Gamma(3-n/2) \times\nonumber \\&&\quad
\Bigl[  S_{Tri}^n(s_2,s_3,s_1,m_2^2,m_3^2,m_1^2) 
       +S_{Tri}^n(s_3,s_1,s_2,m_3^2,m_1^2,m_2^2) 
      + S_{Tri}^n(s_1,s_2,s_3,m_1^2,m_2^2,m_3^2) \Bigr]\nonumber\\&&
\eea
\bea
&& I_4^{n+2}(s_{12},s_{23},s_1,s_2,s_3,s_4,m_1^2,m_2^2,m_3^2,m_4^2) = 
\Gamma(3-n/2) \times\nonumber \\
&&\quad \Bigl[ S_{Box}^{n+2}(s_{23},s_{12},s_{2},s_{3},s_{4},s_{1},m_2^2,m_3^2,m_4^2,m_1^2)  
              +S_{Box}^{n+2}(s_{12},s_{23},s_{3},s_{4},s_{1},s_{2},m_3^2,m_4^2,m_1^2,m_2^2) \nonumber\\
&&\quad       +S_{Box}^{n+2}(s_{23},s_{12},s_{4},s_{1},s_{2},s_{3},m_4^2,m_1^2,m_2^2,m_3^2) 
              +S_{Box}^{n+2}(s_{12},s_{23},s_{1},s_{2},s_{3},s_{4},m_1^2,m_2^2,m_3^2,m_4^2) \Bigr]
	      \nonumber\\
\eea
If no infrared divergences are present one can set $n=4$ in both formulas.
Explicit formulas for IR-finite integrals in $n=4$ dimensions are known for  
triangles and boxes and can be found in
\cite{Denner:qq,vanOldenborgh:1989wn}. For the case of vanishing
internal masses a list of box and triangle integrals may be found in 
\cite{BernDixonKosower,Campbell:1996zw,Binoth:2001vm}.
To the best of our knowledge no complete list of all mixed cases is given
in the literature.
We note in this respect that our representation of 4-dimensional box integrals
in terms of triangle functions and 6--dimensional boxes is a good starting
point, as the infrared divergences are exclusively contained in 
triangle functions in this case.

In both integrals, (\ref{Stri}) and (\ref{Sbox}), the exponent of the kinematic
functions is $-1$. Note that high negative powers of kinematic
functions are typically the reason for the failure of a stable numerical
evaluation of multi-leg or -loop Feynman diagrams in the physical region. 
In this respect our representation is well suited for a numerical approach. 
As the kinematic functions are quadratic in each parameter, 
one integration can be done analytically. For the triangle one gets
\bea\label{Stri_num}
S_{Tri}^{n=4}(s_1,s_2,s_3,m_1^2,m_2^2,m_3^2) 
= \int\limits_0^1 dt_1dt_2 
\frac{1}{(1+t_{1}+t_2)}\frac{1}{A t_2^2 + B t_2 + C - i \delta}\nonumber\\
= \int\limits_0^1 dt_1  \frac{2A}{\sqrt{R}} 
\left[ \frac{\log(2A+B-\sqrt{R}) - \log(B-\sqrt{R}) 
           - \log(2A+B+T)+\log(B+T)}{T+\sqrt{R}} \right.\nonumber\\
-
\left. \frac{\log(2A+B+\sqrt{R}) - \log(B+\sqrt{R}) 
           - \log(2A+B+T)+\log(B+T)}{T-\sqrt{R}}
\right]
\eea 
with
\bea
A &=& m_2^2 \nonumber\\
B &=& (m_1^2+m_2^2-s_2) t_1 + m_2^2 + m_3^2 -s_3 \nonumber\\
C &=& m_1^2 t_1^2  + ( m_1^2+m_3^2-s_1) t_1 + m_3^2 \nonumber\\
R &=& B^2 - 4 A C + i \delta\nonumber\\
T &=& 2 A ( 1+t_1 ) - B
\eea
For the 6--dimensional box integral one finds similarly
\bea\label{Sbox_num}
&&S_{Box}^{n=6}(s_{12},s_{23},s_1,s_2,s_3,s_4,m_1^2,m_2^2,m_3^2,m_4^2) 
= \int\limits_0^1 dt_1dt_2dt_3
\frac{1}{(1+t_{1}+t_2+t_3)^{2}}\frac{1}{A t_2^2 + B t_2 + C - i \delta}\nonumber\\
&&=\int\limits_0^1 dt_1dt_2  \frac{4A^2}{\sqrt{R}}
\Biggl\{
 \frac{4A\sqrt{R}}{(2A+B+T)(B+T)(T^2-R) } \nonumber\\
&&+ \frac{1}{(T+\sqrt{R})^2}
 \Bigl[\log(2A+B-\sqrt{R})-\log(B-\sqrt{R})
  - \log(2A+B+T)+\log(B+T)\Bigr] \nonumber\\
&&- \frac{1}{(T-\sqrt{R})^2}\Bigl[\log(2A+B+\sqrt{R}) - \log(B+\sqrt{R}) 
          - \log(2A+B+T)+\log(B+T) \Bigr]\Biggr\}
\eea     
with\footnote{We do not introduce new symbols $A,B,C,R,T$ for the box. 
Which definition applies is always clear from the context.}
\bea
A &=& m_3^2 \nonumber\\
B &=& (m_1^2+m_3^2-s_{23}) t_1 + (m_2^2+m_3^2-s_{3}) t_2 + m_3^2 + m_4^2 -s_4 \nonumber\\
C &=& (-s_{12}) t_2 + (-s_{1}) t_1 + (-s_2) t_1 t_2 + (1+t_1+t_2)( m_4^2 + m_1^2 t_1 + m_2^2 t_2) 
\nonumber\\
R &=& B^2 - 4 A C + i \delta\nonumber\\
T &=& 2 A ( 1+t_1+t_2 ) - B
\eea
The critical points for the numerical evaluation of both integrals
are vanishing denominators and logarithms with arguments tending to zero.
Before analysing the singularity structure of the
integrands we explicitly separate imaginary and real part.
It is useful to distinguish the cases $R>0$ and $R<0$ in this respect.
If $R<0$, then $C>0$ and $A+B+C>0$, no imaginary part is present.
We find
\bea\label{Stri_final}
&&S_{Tri}^{n=4}(s_1,s_2,s_3,m_1^2,m_2^2,m_3^2) 
= \int\limits_0^1 dt_1  \frac{4 A}{T^2-R} \nonumber\\&&
\Bigl\{
\Bigl[
\log(2A+B+T) - \log(B+T)
\Bigr] 
+\theta(R<0)
\Bigl[ 
       \frac{\log(C) - \log(A+B+C)}{2}  \nonumber\\&&
+\frac{T}{\sqrt{-R}}
\Bigl(  
   \arctan\left(  \frac{\sqrt{-R}}{B} \right) 
  -\arctan\left(  \frac{\sqrt{-R}}{2 A+B}\right) + \pi\;\theta(B < 0 < 2A+B)  
\Bigr)  
\Bigr] \nonumber\\&&
+\theta(R>0)
\Bigl[ 
\frac{T - \sqrt{R}}{2\,\sqrt{R}}
\Bigl(
 \log\left( | 2 A + B - \sqrt{R} |\right) 
-\log\left( | B - \sqrt{R} |\right) -i \pi \theta( B < \sqrt{R}< 2A+B)
\Bigr)\nonumber\\&&
- \frac{T + \sqrt{R}}{2\,\sqrt{R}}
\Bigl(
 \log\left( | 2 A + B + \sqrt{R} |\right) 
-\log\left( | B + \sqrt{R} |\right) +i \pi \theta( B < -\sqrt{R}< 2A+B)
\Bigr)
\Bigr]
\Bigr\}\nonumber\\&&
\eea 
 and
\bea\label{Sbox_final}
&&S_{Box}^{n=6}(s_{12},s_{23},s_1,s_2,s_3,s_4,m_1^2,m_2^2,m_3^2,m_4^2) 
= \int\limits_0^1 dt_1 dt_2 \frac{16 A^2}{(T^2-R)^2} \nonumber\\&&
\Bigl\{ 
\frac{A(T^2-R)}{(2A+B+T)(B+T)}
+ T \, [ \log(2A+B+T) - \log(B+T)  ] \nonumber\\&&
+\theta(R<0)
\Bigl[ 
      T \,\frac{\log(C) - \log(A+B+C)}{2} \nonumber\\&&
+\frac{ T^2 + R}{2\,\sqrt{-R}}
\Bigl(  
   \arctan\left(  \frac{\sqrt{-R}}{B} \right) 
  -\arctan\left(  \frac{\sqrt{-R}}{2 A+B}\right) + \pi\;\theta(B < 0 < 2A+B)  
\Bigr)  
\Bigr] \nonumber\\&&
+\theta(R>0)
\Bigl[ 
  \frac{(T - \sqrt{R})^2}{4\,\sqrt{R}}
\Bigl(
 \log\left( | 2 A + B - \sqrt{R} |\right) 
-\log\left( | B - \sqrt{R} |\right) - i\pi\;\theta( B < \sqrt{R}< 2A+B)
\Bigr)\nonumber\\&&
- \frac{( T + \sqrt{R})^2}{4\,\sqrt{R}}
\Bigl(
 \log\left( | 2 A + B + \sqrt{R} |\right) 
-\log\left( | B + \sqrt{R} |\right) + i\pi\;\theta( B < -\sqrt{R}< 2A+B)
\Bigr)
\Bigr]
\Bigr\}\nonumber\\&&
\eea 
We define the step function $\theta$ to be 1 if its argument is true, and
0 else.
Let us now investigate the singularity structure of the integrands.
At first sight, dangerous, possibly singular denominators are present.
First note that the limit $T\rightarrow 0$ with simultaneously $R\to 0$ 
is unproblematic since $B\geq 2A >0$ and $C\geq A >0$ in this case. 
The integrands then 
are bounded and positive definite, as can be seen by looking
at the starting expressions in (\ref{Stri_num}) and (\ref{Sbox_num}).
Further, a short calculation shows that the limits
$T\rightarrow \pm\sqrt{R}$ with $R>0$ are finite, so 
the integrand is also non-singular in this limit.
The last denominator which can vanish is $\sqrt{\pm R}$. In the limit
$\pm R\rightarrow 0$ the  integrands in (\ref{Stri_final}) and 
(\ref{Sbox_final}) behave as
\bea\label{behaviourRnegReg1}
\sim \pi \,\theta(B<0<2A+B)  
\Bigl[ 
\frac{\theta(R<0)}{\sqrt{-R}} + i\, \frac{\theta(R>0)}{\sqrt{R}}
\Bigr]\quad +\mbox{ finite } 
\eea 
As $R$ is a quadratic form in the integration variables, we have an
integrable singularity of square-root type.
The integrand also exhibits logarithmic singular behaviour
whenever an argument of a logarithm goes to zero, i.e.
at the boundaries of the regions where the integrand develops an imaginary part. 
Hence, it is necessary to have $R\geq 0$ in order to produce a logarithmic
singularity.

\begin{figure}
\begin{center}
\includegraphics[width=\linewidth]{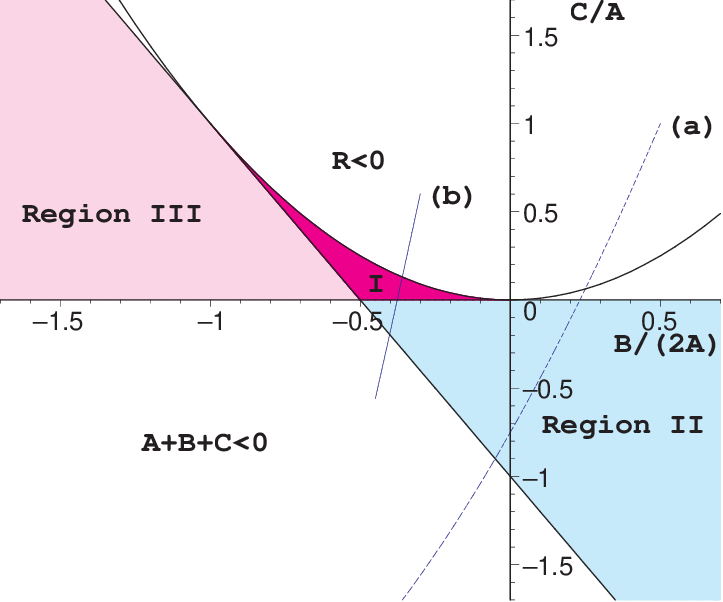}
\end{center}
\caption{\label{Fig:1}{\em Analytic regions of the box and
triangle integrands. The parabola defines the boundary $R=0$,
the line the boundary $A+B+C=0$. Inside regions I,II and III the integrand
has an imaginary part. The integrable square-root and
logarithmic singularities are located at the boundaries of these 
regions as explained in the text. The line segments (a) and (b) stand
for the integration regions of Figs. 2 and 3, respectively.}}
\end{figure}

\begin{figure}[h]
\begin{center}
\includegraphics[width=\linewidth]{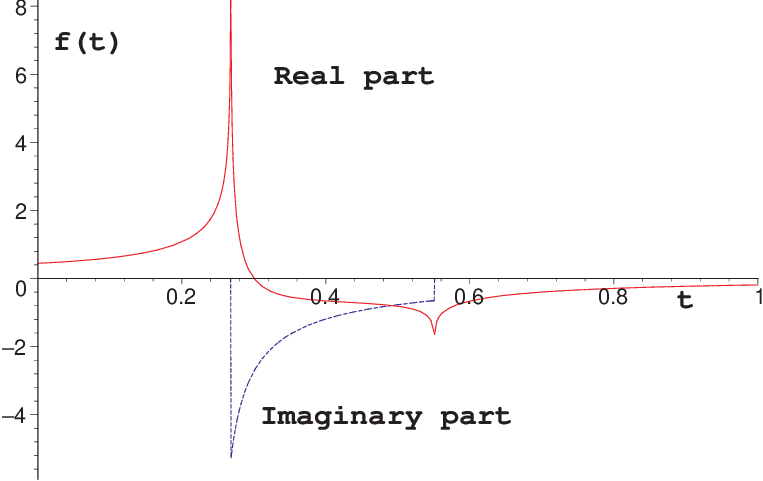}
\end{center}
\caption{\label{Fig:2}{\em The integrand of the function
$S_{Tri}^{n=4}(6,4,1,1,1,1)$} is plotted for $t\in [0,1]$.
The structure of the shown integrand is explained in the text.}
\end{figure}

\begin{figure}
\begin{center}
\includegraphics[width=\linewidth]{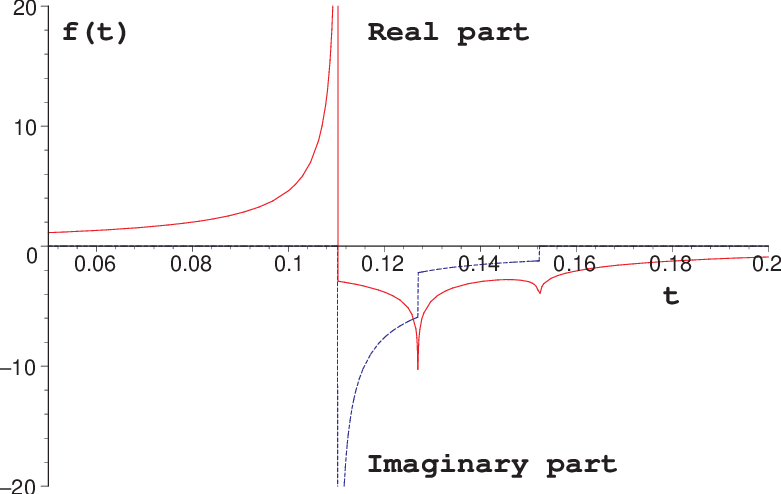}
\end{center}
\caption{\label{Fig:3}{\em The integrand of the function
$S_{Tri}^{n=4}(10,4,5/2,1,1,1)$ is plotted for $t\in [0.05,0.2]$.
Integrable square-root and logarithmic singularities
are visible as explained in the text.}}
\end{figure}

Three regions which lead to an imaginary part can be distinguished:
\begin{description}
\item[Region I:] $A+B+C>0, -2A<B<0, C>0 \Leftrightarrow 
                 ( B<\pm \sqrt{R}<2A+B )$.
\item[Region II:] $A+B+C>0, C<0 \Leftrightarrow 
               (B< \sqrt{R}<2A+B)\, \mbox{and not} \,(B<-\sqrt{R}<2A+B)$.
\item[Region III:]  $A+B+C<0, C>0 \Leftrightarrow  
               (B<-\sqrt{R}<2A+B)\, \mbox{and not} \,(B<\sqrt{R}<2A+B)$.
\end{description}  
Region I is an overlap region where the imaginary part 
has two contributions. In regions II and III only one of the 
$\theta$--functions in (\ref{Stri_final}) and (\ref{Sbox_final}) contributes.
All critical regions are shown in Fig.~\ref{Fig:1},
which illustrates the analytic structure of both integrands.
A given kinematic configuration defines a certain
compact subset in the depicted $C/A,B/(2A)$--plane when the integration
variables are varied from 0 to 1. In the case of the triangle 
the curve
\bea
{\cal C}_{Tri}: t_1 \rightarrow [B(t_1),C(t_1)] &,& t_1 \in [0,1]
\eea
represents the integration contour for a given kinematic configuration,
i.e. a segment of a parabola. In the box 
case one has
\bea
{\cal C}_{Box}:(t_1,t_2)  \rightarrow [B(t_1,t_2),C(t_1,t_2)] &,& t_j \in [0,1] \; ,
\eea
a family of parabolas. The covering of the corresponding region
is multi-valued in general. More precisely, $B$ is linear in the
integration variables and $C$ quadratic. If thresholds are crossed,
$C$ typically goes through a minimum and the integration domain
is two-valued. If the domains of ${\cal C}_{Tri}$ and ${\cal C}_{Box}$
for a given kinematic configuration do not intersect with a boundary of 
the regions I,II and III, the integrand is bounded and no problems
arise for a numerical evaluation. 
For example, in the Euclidean region, where all Mandelstam 
variables are negative, one has always 
$B>0$ and $C>0$ and numerical integration is trivial.
An advantageous feature of the sector decomposition
(\ref{sectordeco}) is that it provides bounded integrands in the 
Euclidean region.  For physical
processes the domains of ${\cal C}_{Tri}$ and ${\cal C}_{Box}$
hit the singularities. 
We display two typical integrands of the triangle graph
for a generic kinematic configuration in Figs.~\ref{Fig:2} and \ref{Fig:3}.
In Fig.~\ref{Fig:2} a situation is shown where
only logarithmically singular behaviour is present. 
The imaginary 
part remains finite in this case. In Fig.~\ref{Fig:3}
one sees the square-root singularities when going from
region $R<0$ to region I. 
The real part diverges
as $R\rightarrow 0$ from below. The imaginary part 
diverges as $R\rightarrow 0$ from above. This is exactly
the  behaviour expected from eq.~(\ref{behaviourRnegReg1}) near this boundary.
The two logarithmic cusps are located at the boundaries of 
region II. Going from region I to region II leads
to a step in the imaginary part. Similar plots are obtained by looking
at the crossing through regions I and III. In the case of 
$S^{n=6}_{box}$ one has a family of such curves labelled by 
one of the integration variables.     

A suitable numerical integration routine has
to succeed in the critical regions. In principle
our formulas contain enough information to explicitly
detect and classify all possible cases and to make
an adequate variable transformation at each critical point. 
Another possibility is to
subtract singular approximations to the integral,
explicitly integrate the singular parts and add the corresponding
value to the result. In both
ways smooth and bounded integral representations can be achieved 
that are free from numerical problems. However, since we have found
an iterative numerical method that {\it automatically} copes with the present
integrable singularities, we do not pursue these approaches further
in this paper.

We want to close this section by remarking
that we implicitly gave a constructive proof of the
fact that {\em any} massive one-loop Feynman diagram
has at most integrable singularities of square-root 
or logarithmic type. Further we point out 
that all the steps of our derivation go through in 
the case of infrared singularities, i.e. in the presence of massless
propagators, as the reduction formulas are also valid in the massless case
and after reduction the IR singularities are isolated in the form of 
triangle functions. 
The necessary modifications of the given formulas for the remaining infrared 
finite parts are straightforward.

\section{Numerical evaluation}

In the previous sections we showed that all 
finite scalar $N$-point functions
can be expressed as linear combinations of
the ``atomic'' integrals defined in (\ref{Stri_final}) and (\ref{Sbox_final}).
In this section we discuss two methods that allow to integrate these
building blocks numerically without the need for further analytic manipulations.

The characteristics of these elementary integrals were discussed in detail
above.  In the current context we recall first that the integration region is a
simple one, namely the unit interval or unit square.  In this case the
optimal approach to achieve high precision rapidly is to use deterministic
integration rules, particularly for low-dimensional, well-behaved integrals.
The case at hand, however, is -- as also discussed above -- complicated by the
possible presence of multiple discontinuities or integrable singularities.  A
well suited strategy for these badly-behaved integrands is to divide the
integration region into subregions, and apply an iterative, adaptive algorithm.
This approach has been used successfully for irregular high-dimensional
integrands in combination with Monte Carlo methods \cite{VEGAS}, but has also
been explored for lower-dimensional integrands in connection with integration
rules \cite{GenzEspelidLyness}.

Since the integrand of the triangle function (\ref{Stri_final}) is only
1-dimensional, we can rely in this case on the efficient and highly robust
QAGS routine of QUADPACK \cite{QUADPACK}, a widely-used package for the numerical
computation of definite 1-dimensional integrals.  The QAGS algorithm applies
an integration rule adaptively in subintervals until
the error estimate is sufficiently small. The results are extrapolated using
the epsilon-algorithm, which accelerates the convergence of the integral in
the presence of discontinuities and integrable singularities. The maximal
number of subintervals is a fixed input parameter.  A maximum of 1000
subintervals proved sufficient to achieve the desired relative error of
$10^{-4}$ in our sample calculations.  With this choice, the runtime 
consumed by triangle evaluations turned out to be negligible compared to the 
one for the box evaluations.

As succinctly discussed in \cite{NumericalRecipes}, Section 4.6, evaluating
multi-dimensional integrals like function (\ref{Sbox_final}) poses more
difficulties. On the other hand, since the box integral is only 2-dimensional, 
a generalization of the efficient, deterministic method we selected to evaluate the
1-dimensional triangle function is suggestive, and several algorithms have
indeed been discussed in the literature \cite{GenzEspelidLyness}.  Their
application, however, would require analytic knowledge of the location of all
singularities.  Hence these algorithms cannot be applied directly to the
integral in (\ref{Sbox_final})\footnote{As mentioned above, it is possible
to determine analytically the location of the singularities of the box
function, so in principle, an entirely deterministic integration of the box
function is also feasible.}.  We therefore proceed by decomposing the potentially
singular integrand $f(x)$ into a bounded function $b_c(x)$ and a singular rest
$s_c(x)$ by introducing a cut parameter $c > 0$:
\be
b_c(x) :=
  \left\{
  \begin{array}{lcl}
  +c & \mbox{if} & f(x) > c \\
  f(x) & \mbox{if} & -c \leq f(x) \leq c \\
  -c & \mbox{if} & f(x) < -c
  \end{array}
  \right.
\ee
\be
  s_c(x) := f(x) - b_c(x)
\ee

Relying again on integration rules for maximum efficiency, we can then
integrate $b_c(x)$ with DCUHRE \cite{DCUHRE}, a robust, globally adaptive
algorithm applicable to multidimensional, bounded integrands.  To integrate
$s_c(x)$ we revert to a well-established, non-deterministic alternative that
requires no detailed knowledge of the singularity structure of the integrand:
Monte Carlo integration \cite{NumericalRecipes}.  This advantage, however, is
offset by a significantly slower convergence relative to deterministic
methods.  This interplay suggests the existence of an optimal range for the
cut parameter $c$.  Below that range, unnecessarily large, non-singular
regions are Monte-Carlo integrated causing slow convergence.  Above that
range, the deterministic routine has to integrate unnecessarily steep peaks
in the singular regions, necessitating a large number of subdivisions and
function evaluations and consequently a large workspace due to the global
nature of its algorithm.  On our systems, the workspace limit was about~350MB,
allowing for a maximum of $1.5\cdot10^9$ function evaluations for DCUHRE.
For the calculations shown in the next section we experimented with values
for $c$ from 500 to 50000. The optimal range was 5000 to 10000.  The final
result is obtained by adding up the integrals over $b_c(x)$ and $s_c(x)$.  As
expected, one finds that the obtained results are independent of the
parameter $c$.

We employed an optimised approach to the Monte Carlo integration of $s_c(x)$
that requires a 2-dimensional grid covering the integration region.  During
the integration of $b_c(x)$, some evaluations of $f(x)$ may return values above
or below the cut, i.e. $s_c(x) \neq 0$.  In this case the grid cell that contains
$x$ is saved.  In a second step the integral over $s_c(x)$ is calculated by using
all cells with $s_c(x) \neq 0$ detected in the previous step as seed cells.
$s_c(x)$ is integrated in each cell using crude Monte Carlo integration.  If a
cell result is finite, all neighboring cells are also evaluated.  This procedure
is applied recursively until the region where $s_c(x)$ is non-vanishing is covered.

Due to its global nature the method described so far requires a potentially large
amount of memory, and the question arises if a viable ``local'' alternative with
negligible memory requirements can be found.  To that end, we propose a second,
fully recursive approach. Assume the integral $I_0$ over a hypercube with
volume $V_0$ is to be determined with precision $\Delta I_0$.  Starting with
volume $V_0$ the following procedure is applied recursively:
\begin{enumerate}
\item A value $I$ and error estimate $\Delta I$ for the integral in the cell of
volume $V$ is obtained by applying
   \begin{itemize}
   \item[a)] an integration rule (ca. 200 integrand evaluations are necessary for a
             degree 13 integration rule \cite{DCUHRE})
   \item[b)] basic Monte Carlo integration with the same number of function
             evaluations as in a)
   \end{itemize}
   If both results are compatible within errors, the one with the lower error
   estimate is selected, otherwise the result with the larger error is selected.
\item The tolerable error\footnote{This condition guarantees that the overall error
      is at most $\Delta I_0$.} in the cell is
      $\Delta I_{max} := \Delta I_0\sqrt{V/V_0}$.
\item If $\Delta I \leq \Delta I_{max}$ no further action is necessary.
      If $\Delta I > \Delta I_{max}$ the cell is divided into $n$ subcells of equal
      volume, and the integrals $I_i$ in the subcells are determined as in 1.
      
\item If $\Delta I_{div} < \Delta I/\sqrt{n}$ with $ \Delta I_{div} :=
      \left[\sum_{i=1}^n (\Delta I_i)^2\right]^{1/2}$, the procedure is applied
      recursively to the subcells.
\item If $\Delta I_{div} \geq \Delta I/\sqrt{n}$, further subdivision is not
      advantageous, and $I$ is Monte Carlo sampled until
      $\Delta I \leq \Delta I_{max}$.
\end{enumerate}
This algorithm does not involve a cutoff parameter and was used to double-check
the results obtained with the first method.  For typical box functions with
singularities we observe no clear superiority of integration rule versus Monte
Carlo evaluations, and the added flexibility indeed increases efficiency.
To the best of our knowledge, the combined application of integration rules and
Monte Carlo sampling has not previously been proposed in the literature.

Having error estimates for the elementary triangle and box integrals, error
estimates for the scalar 4-, 5-, and 6-point functions are obtained using
standard error propagation.  The estimated relative error of the results
presented in Tables \ref{table1} and \ref{table2} is $10^{-4}$ or better.

To conclude, we note
that the runtime of our program for the scalar hexagon function
depends strongly on the complexity imposed by the kinematic configuration
and ranges from less than 30 seconds to many hours. 
For lower precision the runtime improves considerably.

\section{Results}
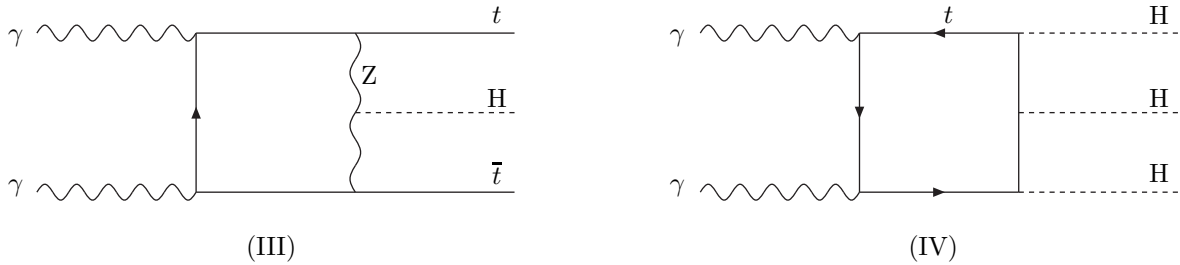
\begin{figure}[htb]
\begin{picture}(100,17)(-15,-20)
\Photon(-60,30)(0,30){3}{5}
\Photon(-60,-30)(0,-30){3}{5}
\ArrowLine(0,-30)(0,30)
\Line(0,30)(120,30)
\Line(0,-30)(120,-30)
\Photon(60,30)(60,-30){2}{3}
\DashLine(60,0)(120,0){2}
\Text(-24,10)[c]{$\gamma$}
\Text(-24,-10)[c]{$\gamma$}
\Text(40,2)[c]{H}
\Text(23,5)[c]{Z}
\Text(40,13)[c]{$t$}
\Text(40,-8)[c]{$\overline{t}$}
\Text(10,-18)[c]{(III)}
\Photon(190,30)(250,30){3}{5}
\Photon(190,-30)(250,-30){3}{5}
\ArrowLine(250,30)(250,-30)
\ArrowLine(310,30)(250,30)
\ArrowLine(250,-30)(310,-30)
\Line(310,-30)(310,30)
\DashLine(310,30)(370,30){2}
\DashLine(310,0)(370,0){2}
\DashLine(310,-30)(370,-30){2}
\Text(64,10)[c]{$\gamma$}
\Text(64,-10)[c]{$\gamma$}
\Text(128,2)[c]{H}
\Text(100,13)[c]{$t$}
\Text(128,13)[c]{H}
\Text(128,-8)[c]{H}
\Text(98,-18)[c]{(IV)}
\end{picture}
\caption{The diagrams containing the scalar integrals III and IV used as benchmarks in Table \ref{table1}.}
\label{penta}
\end{figure}

To demonstrate the practicality of the approach described above, we calculate
the 4-dimensional scalar pentagon and hexagon functions for several physical
and unphysical kinematic configurations. 
In Table~\ref{table1} we give  numerical results for the pentagon 
integral. Values are shown for two Euclidean (I and II) 
and two physical kinematic configurations. 
\begin{table}[htb]
\begin{center}
\begin{tabular}{|l|r|r|r|r|}
\hline
 & I & II & III & IV \\
\hline
$s_{12}$ & -1 &   -1 &       4 &      4 \\
$s_{23}$ & -1 &   -1 &    -1/5 &  -7/10 \\
$s_{34}$ & -1 &   -1 &     1/5 &   1/10 \\
$s_{45}$ & -1 &   -1 &    3/10 &   3/10 \\
$s_{51}$ & -1 &   -1 &    -1/2 &   -1/2 \\
$s_1$ &    -1 &   -1 &       0 &      0 \\
$s_2$ &    -1 &   -1 &       0 &      0 \\
$s_3$ &    -1 &   -1 &  49/256 &  9/100 \\
$s_4$ &    -1 &   -1 &   9/100 &  9/100 \\
$s_5$ &    -1 & -5/2 &  49/256 &  9/100 \\
$m_1^2$ &   1 &    1 &  49/256 & 49/256 \\
$m_2^2$ &   1 &    1 &  49/256 & 49/256 \\
$m_3^2$ &   1 &    1 & 81/1600 & 49/256 \\
$m_4^2$ &   1 &    1 & 81/1600 & 49/256 \\
$m_5^2$ &   1 &    1 &  49/256 & 49/256 \\
\hline
Re & -0.03542 & -0.03203 &  41.33 & 3.533 \\
Im &        0 &        0 & -45.96 & -5.956 \\
\hline
\end{tabular}
\caption{\label{table1}Four-dimensional scalar pentagon function evaluated for 
unphysical (I and II) and physical (III and IV) kinematics. The diagrams defining
the kinematics III and IV are depicted in Fig.~\ref{penta}.
All energies and masses are scaled by $E_{cms}/2 =$ 400 GeV.}
\end{center}
\end{table}
The latter correspond to scalar integrals occurring in the computation
of the process $\gamma\gamma \rightarrow t \bar t H$ (pentagon III),
and  $\gamma\gamma \rightarrow H H H$ (pentagon IV), see Fig.~\ref{penta}.
To work with realistic mass scales we use $E_{cms}=800$ GeV,
$m_Z= 90$ GeV, $m_{top}= 175$ GeV, $m_{Higgs}= 120$ GeV.
All kinematic invariants are rescaled by $(E_{cms}/2)^2$.

\begin{figure}[htb]\label{hexa}
\begin{picture}(100,32)(-15,-22)
\Photon(-60,30)(0,30){3}{5}
\Photon(-60,-30)(0,-30){3}{5}
\ArrowLine(0,-30)(0,30)
\Line(0,30)(120,30)
\Line(0,-30)(120,-30)
\Photon(60,30)(60,-30){2}{3}
\Photon(60,0)(120,0){2}{3}
\DashLine(30,-30)(90,-60){2}
\Text(-24,10)[c]{$\gamma$}
\Text(-24,-10)[c]{$\gamma$}
\Text(40,2)[c]{Z}
\Text(23,5)[c]{Z}
\Text(40,13)[c]{$t$}
\Text(40,-8)[c]{$\overline{t}$}
\Text(35,-23)[c]{H}
\Text(10,-18)[c]{(II)}
\Photon(190,30)(250,30){3}{5}
\Photon(190,-30)(250,-30){3}{5}
\ArrowLine(250,30)(250,-30)
\Line(310,30)(250,30)
\Line(250,-30)(310,-30)
\ArrowLine(310,-30)(310,30)
\DashLine(310,30)(370,30){2}
\DashLine(310,-30)(370,-30){2}
\DashLine(280,30)(340,60){2}
\DashLine(280,-30)(340,-60){2}
\Text(64,10)[c]{$\gamma$}
\Text(64,-10)[c]{$\gamma$}
\Text(117,23)[c]{H}
\Text(117,-23)[c]{H}
\Text(100,8)[c]{$t$}
\Text(128,13)[c]{H}
\Text(128,-13)[c]{H}
\Text(98,-18)[c]{(III)}
\end{picture}
\caption{The graphs II and III defining the kinematics for the scalar integrals 
calculated in Table~\ref{table2}.}
\end{figure}
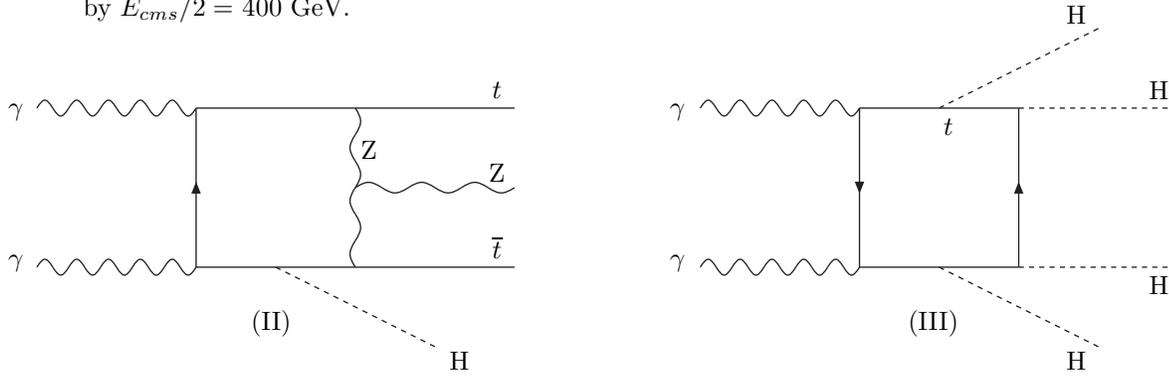

\clearpage
In Table~\ref{table2} we display the numerical result for the 
hexagon integral at the
modified\footnote{
The symmetric point for the hexagon function does not obey the 
nonlinear constraint $\det G=0$,
where $G$ is the Gram matrix of the six--point kinematics.
The {\it modified} symmetric point is designed to 
satisfy the constraint $\det G=0$, which determines the value for $s_{345}$.} 
symmetric point (I) and  two physical points (II and III).
Point II corresponds to a kinematic situation arising in the
process $\gamma\gamma \rightarrow t \bar t H Z$ and 
point III  to a kinematic situation in the
process $\gamma\gamma \rightarrow H H H H$  (see Fig.~\ref{hexa}). 
\begin{table}[htb]
\begin{center}
\begin{tabular}{|l|r|r|r|}
\hline
 & I & II & III \\
\hline
$s_{12}$ &    -1 &           4 &          4 \\
$s_{23}$ &    -1 &       -1/10 &       -1/5 \\
$s_{34}$ &    -1 &         1/5 &        1/5 \\
$s_{45}$ &    -1 &        3/10 &        2/5 \\
$s_{56}$ &    -1 &         2/5 &       3/10 \\
$s_{61}$ &    -1 &        -1/5 &      -1/10 \\
$s_{123}$ &   -1 &        3/10 &       1/10 \\
$s_{234}$ &   -1 &        -1/5 &      -3/10 \\
$s_{345}$ & -5/2 & 1.753247474 & 0.38189943 \\
$s_1$ &       -1 &           0 &          0 \\
$s_2$ &       -1 &           0 &          0 \\
$s_3$ &       -1 &      49/256 &      9/100 \\
$s_4$ &       -1 &     81/1600 &      9/100 \\
$s_5$ &       -1 &      49/256 &      9/100 \\
$s_6$ &       -1 &       9/100 &      9/100 \\
$m_1^2$ &      1 &      49/256 &     49/256 \\
$m_2^2$ &      1 &      49/256 &     49/256 \\
$m_3^2$ &      1 &     81/1600 &     49/256 \\
$m_4^2$ &      1 &     81/1600 &     49/256 \\
$m_5^2$ &      1 &      49/256 &     49/256 \\
$m_6^2$ &      1 &      49/256 &     49/256 \\
\hline
Re & 0.01353 & -653.8 & -26.93 \\
Im &       0 &  3.24 &  48.63 \\
\hline
\end{tabular}
\caption{Four-dimensional scalar hexagon function evaluated for unphysical (I) and
physical  (II and III) kinematics. The
corresponding diagrams are shown in Fig.~\ref{hexa}. 
All energies and masses are scaled by $E_{cms}/2 =$ 400 GeV.\label{table2}}
\end{center}
\end{table}

\clearpage 
The symmetric and modified symmetric points were used to
check our implementation by 
confirming relations between hexagon and pentagon formulas.
To be specific, we 
considered various $N$--point functions at the symmetric Euclidean point:
\begin{eqnarray}
I_{3,sym}^{n=4} & := & I_3^{n=4}(s_1 = s_2 = s_3 = -1, m_1^2 = m_2^2 = m_3^2 = 1)\ \ =\ -0.401140\nonumber\\
I_{4,sym}^{n=6} & := & I_4^{n=6}(s_i = s_{ij} = - m_i^2 = -1)\ \  =\ \  0.128436 \nonumber\\
I_{4,sym}^{n=4} & := & I_4^{n=4}(s_i = s_{ij} = - m_i^2 = -1)\ \  =\ \  0.0991651 \nonumber\\
I_{5,sym}^{n=4} & := & I_5^{n=4}(s_i = s_{ij} = - m_i^2 = -1)\ \  =\  -0.0354161 \nonumber\\
I_{5,sym-mod}^{n=4} & := & I_5^{n=4}(s_5 = -5/2,\: {\rm else}\; s_i = s_{ij} = - m_i^2 = -1)
\ \  =\   -0.0320346 \nonumber\\
I_{6,sym-mod}^{n=4} & := & I_6^{n=4}(s_{345} = -5/2,\: {\rm else}\; s_i = s_{ij} = s_{ijk} = - m_i^2 = -1)\ \  =\ \  0.013526 \nonumber
\end{eqnarray}
For these special kinematic configurations the reduction simplifies 
significantly, and one can show that the following identities hold:
\be
  I_{4,sym}^{n=4} = -\frac{4}{11}\, \left(I_{3,sym}^{n=4} + I_{4,sym}^{n=6}\right)
\ee
\be
  I_{5,sym}^{n=4} = -\frac{5}{14}\, I_{4,sym}^{n=4}
\ee
\be
  I_{6,sym-mod}^{n=4} = -4 \, \left(I_{5,sym}^{n=4} - I_{5,sym-mod}^{n=4}\right)
\ee
Our numerical results -- also shown above -- fulfill all identities.
In addition, we double-checked the results using the 
well-tested program described in~\cite{Binoth:2000ps} 
to calculate multi-loop integrals numerically 
in the Euclidean region.

The threshold behaviour of the box and hexagon representations is probed in
threshold scans, which are shown in Figs.~\ref{ts4} and \ref{ts6}.

\begin{figure}[htb]
\begin{picture}(200,50)(-15,-23)
\Photon(-60,30)(0,30){2}{3}
\Photon(-60,-30)(0,-30){2}{3}
\ArrowLine(0,-30)(0,30)
\Line(0,30)(120,30)
\Line(0,-30)(120,-30)
\Photon(60,30)(60,-30){2}{3}
\Text(-24,10)[c]{Z}
\Text(-24,-10)[c]{Z}
\Text(24,0)[c]{W}
\Text(40,13)[c]{$b$}
\Text(12,13)[c]{$t$}
\Text(40,-13)[c]{$\overline{b}$}
\Text(10,-19)[c]{(a)}
\Photon(190,30)(250,30){3}{5}
\Photon(190,-30)(250,-30){3}{5}
\ArrowLine(250,30)(250,-30)
\Line(310,30)(250,30)
\Line(250,-30)(310,-30)
\ArrowLine(310,-30)(310,30)
\Photon(310,30)(370,30){2}{3}
\Photon(310,-30)(370,-30){2}{3}
\Photon(280,30)(335,55){3}{5}
\Photon(280,-31)(335,-55){3}{5}
\Text(64,10)[c]{$\gamma$}
\Text(64,-10)[c]{$\gamma$}
\Text(117,22)[c]{$\gamma$}
\Text(117,-22)[c]{$\gamma$}
\Text(90,5)[c]{$t$}
\Text(128,13)[c]{Z}
\Text(128,-14)[c]{Z}
\Text(98,-19)[c]{(b)}
\end{picture}
\caption{The box and hexagon diagrams which define the kinematics for the scalar 
integrals used in the
threshold scans shown in Figures \ref{ts4} and \ref{ts6}.}
\label{tscan}
\end{figure}
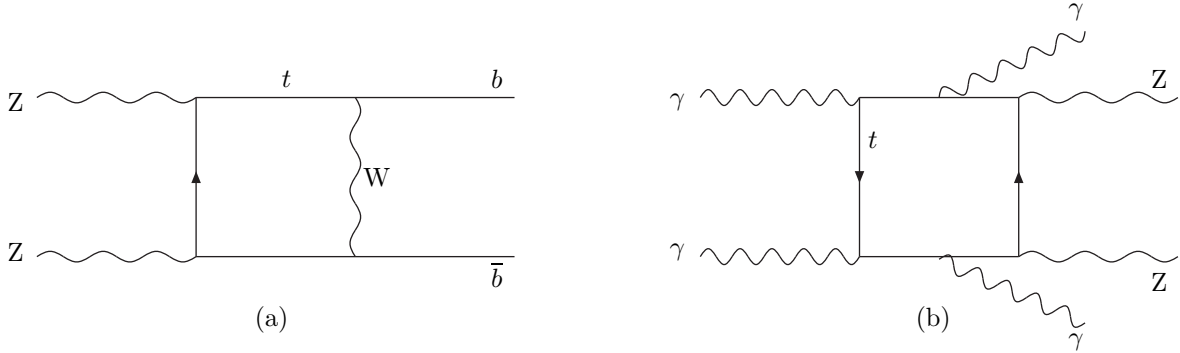

The kinematic configuration for the threshold
scan  of the box function (Figs.~\ref{tscan}a and \ref{ts4}) is
\bea
&&s_{12} = (E_{cms}/(2m_t))^2,\; s_{23} = -(m_Z/(2m_t))^2/2,\nonumber\\
&&s_1 = s_2 = (m_Z/(2m_t))^2,\; s_3 = s_4 = (m_b/(2 m_t))^2, \nonumber\\
&&m_1^2 = m_2^2 =m_4^2 =  1/4,\; m_3^2 = ( m_W/(2m_t) )^2\nonumber
\eea
with $m_b=5$\,GeV and $m_W=80$\,GeV.

The kinematic configuration for the threshold scan of the hexagon function 
(see Figs.~\ref{tscan}b and~\ref{ts6}) 
is defined by 
\bea
&& s_{12} = (E_{cms}/(2m_t))^2,\nonumber\\
&& s_{23} = -s_{34} = s_{61} = -s_{123}
= -1/10,\, s_{45} = -s_{234} = 2/10,\; s_{345} = s_{345}(E_{cms}),\; s_{56} = 3/10,\nonumber\\
&& s_1 = s_2 = s_3 = s_6 = 0,\; s_4 = s_5 = (m_Z/(2m_t))^2,\; m_i^2 = 1/4.
\eea
$E_{cms}$ has been varied between 200 and 600 GeV.
\begin{figure}[htb]
\vspace*{0.5cm}
\begin{center}
\begin{minipage}[c]{.48\linewidth}
\flushright \includegraphics[width=6.5cm, angle=90]{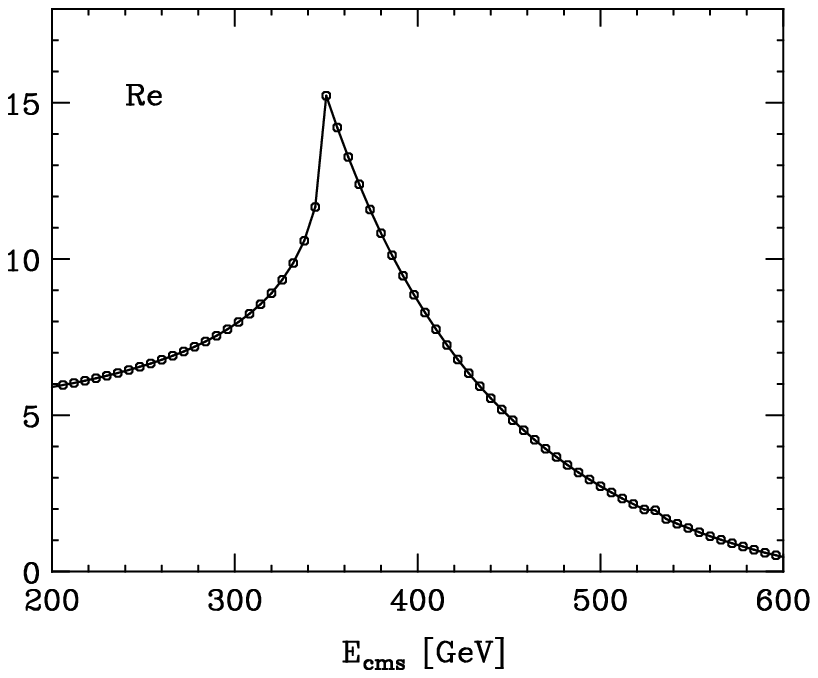} 
\end{minipage} \hfill
\begin{minipage}[c]{.48\linewidth}
\flushleft \includegraphics[width=6.5cm, angle=90]{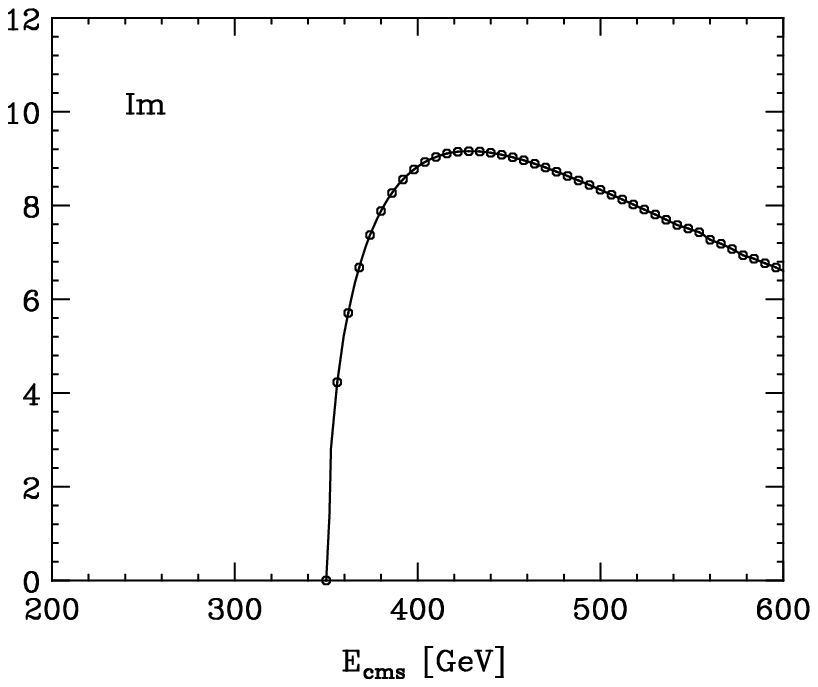}
\end{minipage}
\end{center}
\caption{Scan of the $2m_t = 350$ GeV threshold of the 4-dimensional scalar 
box function, arising from the diagram in Fig.~\ref{tscan}\,(a). Details are 
given in the text.}
\label{ts4}
\end{figure}

\begin{figure}[htb]
\vspace*{0.5cm}
\begin{center}
\begin{minipage}[c]{.48\linewidth}
\flushright \includegraphics[width=6.5cm, angle=90]{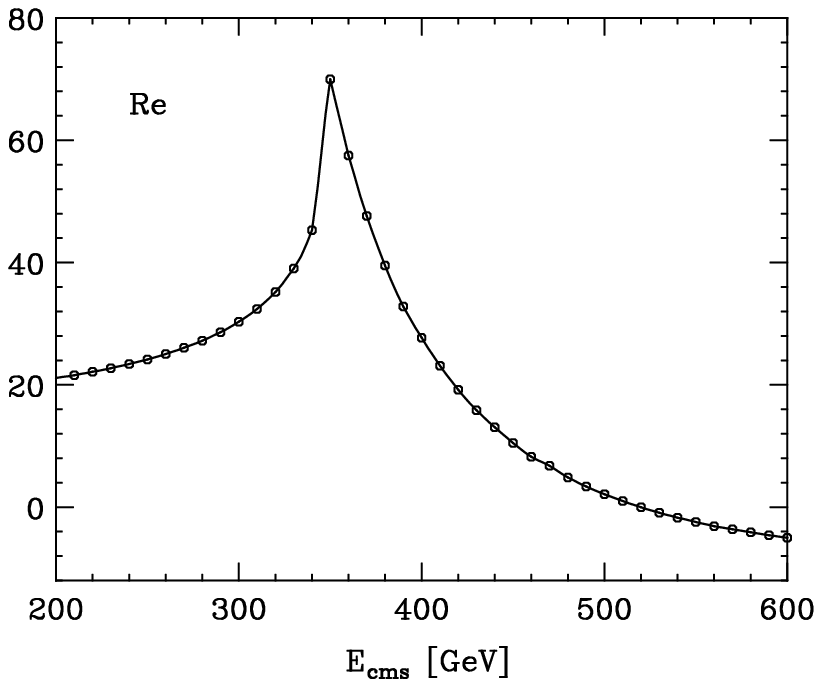} 
\end{minipage} \hfill
\begin{minipage}[c]{.48\linewidth}
\flushleft \includegraphics[width=6.5cm, angle=90]{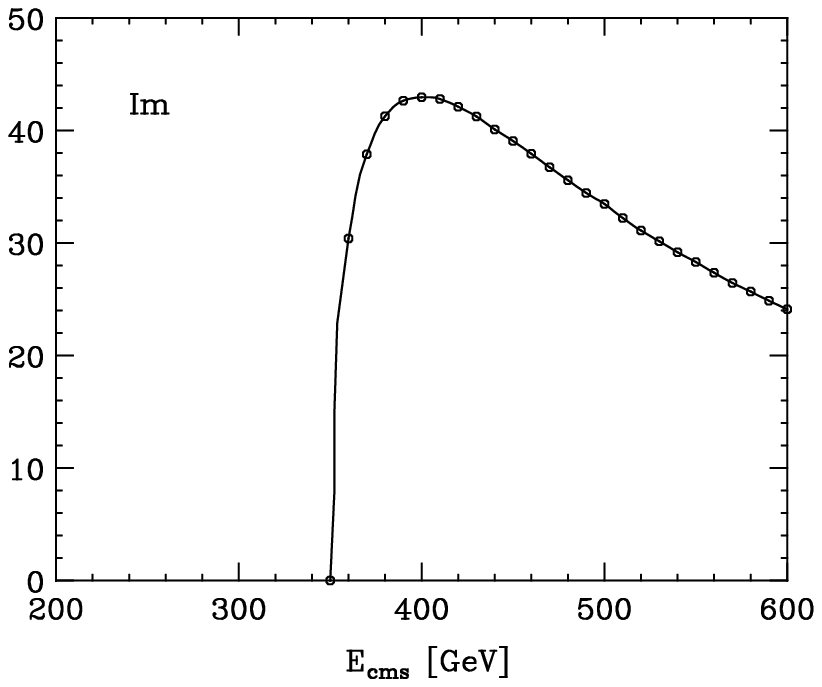}
\end{minipage}
\end{center}
\caption{Scan of the $2m_t = 350$ GeV threshold of the 4-dimensional 
scalar hexagon function, see Fig.~\ref{tscan}b. 
The kinematics is defined in the text.}
\label{ts6}
\end{figure}
We further tested the correctness of the box and pentagon routines by comparing
physical results with imaginary parts against results obtained using the analytic
formulas implemented in\,\cite{vanOldenborgh:1989wn,looptools} and found consistency.

\section{Conclusion}

We have provided explicit representations for the general scalar
box, pentagon and hexagon integrals in terms of n--dimensional
triangle and (n+2)--dimensional box functions. The advantage of such 
a decomposition is that IR divergences, if present, 
can be isolated easily, such that an efficient book-keeping of IR poles
is achieved. 
The remaining integrals can be evaluated by setting $\epsilon=0$.
In our approach, the finite triangles and 6--dimensional box functions 
are represented as one- respectively two--dimensional parameter integrals 
in a form which is convenient for numerical integration.  
Although in principle all these integrals can be evaluated 
analytically by applying scalar reduction formulas, 
it is clear that the expression
for the general hexagon function would contain a huge number of 
dilogarithms which would give rise to nontrivial cancellations.
Representations with a smaller number of ``atoms'' seem to be
preferable from this point of view. 

This motivated us in the present work to 
follow a numerical approach at an earlier stage of the calculation, where
the ``atoms'' 
are the  n--dimensional triangle and the (n+2)--dimensional box function
instead of logarithms and dilogarithms.
Focusing on the massive case we  studied in detail the singularity structure 
of our one (two)--dimensional 
integral representations for the triangle (box)  functions. 
Real and imaginary parts are explicitly separated in this approach 
which avoids to deal with infinitesimal quantities. 
We have shown for physical kinematics that the integral representations
for triangle and 6--dimensional box have an analogous singularity 
structure which can be treated transparently for both in the same way. 

In principle our formulation contains
enough information to produce bounded and smooth integrands
by  adequate variable transformations or subtractions at the critical points. 
However, it seemed an interesting question to us  
whether such somewhat cumbersome manipulations can be avoided by directly 
evaluating the integral representations. To our best knowledge
no general numerical algorithm has been  provided in the literature so far 
for the relatively complicated singularity structure present in the
2--parameter integral representation of the 6--dimensional box.   
By combining deterministic integration methods, adequate for the
smooth part of the integrand, and  Monte Carlo techniques for
the singular regions in an iterative way, we succeeded in
numerically evaluating the integrals directly. Two different 
numerical integration methods have been presented. 
The correctness and efficiency of  our integration strategies 
has been demonstrated
by comparing our results to known results if available.
For the hexagon integral, which could only be tested
indirectly, we have performed several consistency checks. 
We have shown that a stable result can be 
obtained with our method  when internal thresholds are probed. 

A natural question to ask next is if a fully numerical approach
to the evaluation of multi-particle production at 
one-loop is feasible. Analytic calculations are generally hampered by
the enormous complexity generated when reducing
integrals with nontrivial numerators to scalar integrals.
It is conceivable that the critical reduction steps could be avoided
to a large extent if all or some groups
of Feynman diagrams are treated numerically, such that one is 
dispensed from doing the full tensor reduction. 
We note in this respect that the numerical difficulties
stem entirely from the denominators of the integrals, whereas tensor
integrals only introduce additional parameters in the
numerators, which pose no problem. 

Following our strategy
one is always able to separate real and imaginary parts, after
having done the trivial integrations analytically. The remaining
integrals will always have at most singularities of  
square root and/or logarithmic type, which can be integrated directly 
with our methods. 
Thus our findings can be viewed as a step towards a complete numerical approach
to calculate multi--scale processes at one loop. 

\section*{Acknowledgements}
GH would like to thank Fukuko Yuasa for having pointed out the mistake in 
the imaginary part of point II in Table 2,
which has been corrected in the present version. 
The new value has been verified using the program {\sc SecDec}\,\cite{Borowka:2012yc}.

\end{document}